\documentclass{elsart}
\usepackage{amsmath, amssymb}

\begin{document}

\begin{frontmatter}

\title{The general harmonic-oscillator brackets: \\
compact expression, symmetries, sums \\
and Fortran code}
\author [VDU]{G.P. Kamuntavi\v{c}ius},
\author [FI]{R.K. Kalinauskas},
\author [AU]{B.R. Barrett},
\author [VDU]{S. Mickevi\v{c}ius}, and
\author [VDU]{D. Germanas}

\address[VDU]{Vytautas Magnus University, Kaunas LT-3000 Lithuania}
\address[FI]{Institute of Physics, Vilnius LT-2600 Lithuania}
\address[AU]{Department of Physics, University of Arizona, Tucson, Arizona 85721}

\begin{abstract}
We present a very simple expression and a Fortran code for the fast and
precise calculation of three-dimensional harmonic-oscillator transformation
brackets. The complete system of symmetries for the brackets along with
analytical expressions for sums, containing products of two and three
brackets, is given.
\end{abstract}

\end{frontmatter}

\section{Introduction}

A basis of harmonic-oscillator functions has proven to be extremely useful
and efficient in describing compact quantum systems, such as nucleons in
atomic nuclei and quarks in hadrons. The traditional applications, such as
the nuclear shell-model, however, are based on a model Hamiltonian with
individual one-particle variables. Hence, the corresponding model wave
functions, which are dependent on one-particle coordinates, are not
translationally invariant and cannot represent the wave function of such a
system in proper way because the center of mass (c.m.) of a free nucleus (or
a free hadron) must be described by an exponential function, corresponding
to a freely-moving point mass. This need not to be a problem in the case
when the expression for the realistic wave function in an
harmonic-oscillator basis is known. In fact, having this expression, at
least two possibilities exist to find a solution of the problem of
translational invariance of the wave function. The first approach is based
on constructing the superpositions of shell-model states which possess a
fixed state for the c.m. of the system, see \cite{NS69}. Utilizing a basis
of this kind, we do not need to worry about the c.m. state because all
operators for observables are translationally invariant so that the final
result is independent of this state. The second possibility is based on a
direct construction of the many-fermion wave function, which is independent
of the c.m. coordinate, see \cite{VV70}, \cite{GK89}, \cite{NKB}. In this
case, the harmonic-oscillator basis set, in terms of intrinsic (Jacobi)
coordinates, is necessary. By a set of Jacobi coordinates for a system of $N$
particles, we mean the $N-1$ independent vectors that each represents the
displacement of the c.m. of two different subsystems. For $N>2,$ there
exists more than one set of Jacobi coordinates that can be assigned to an $N$%
-particle system. The different possible sets of Jacobi coordinates can be
identified by use of a Jacobi tree (see, for example, \cite{SS77})\ and are
related to each other by orthogonal transformations. In general, when the
transformation from one set of Jacobi coordinates to another is performed,
one obtains an expansion for the wave function containing an infinite number
of terms. Only the set of harmonic-oscillator functions can be chosen in
such a manner that the transformation from one set of Jacobi coordinates to
another results in a corresponding expansion with a finite number of terms.
In any of the above mentioned approaches, the essential feature is the
Talmi-Moshinsky transformation \cite{Talmi}, \cite{Mosh}, \cite{Smirn} and
corresponding harmonic-oscillator brackets (HOBs). Historically, the first
application of this procedure was transformation of the product of two
harmonic-oscillator functions from single-particle coordinates to relative
motion and c.m. coordinates, a reduction that has proven to be very useful
for the evaluation of two-body matrix elements. Because the HOBs are
constantly employed in various model calculations of nuclear and hadrons
structure, it is desirable to have a simple and efficient method for
calculating them. Many papers have been devoted to the study of the
Talmi-Moshinsky transformation and brackets, and various methods for the
calculation of these brackets and several explicit expressions for them are
described in the literature (see \cite{Trlifai} and references therein, \cite%
{BM} and references therein). In this paper we present a complete system of
symmetries for these brackets, a very simple expression for the HOBs based
on the result of \cite{BM}, new expressions for the sums of products of HOBs
and a computer code, written in Fortran, which calculates the HOBs quickly
and precisely.

\section{The definition}

First, let us consider the HOBs, defined in the following way: 
\begin{equation}
\left| e_{1}l_{1}\left( \mathbf{r}_{1}\right) ,e_{2}l_{2}\left( \mathbf{r}%
_{2}\right) :\Lambda \lambda \right\rangle =\sum_{EL,el}\left\langle
EL,el:\Lambda \right. \left| e_{1}l_{1},e_{2}l_{2}:\Lambda \right\rangle
_{d}\left| EL\left( \mathbf{R}\right) ,el\left( \mathbf{r}\right) :\Lambda
\lambda \right\rangle ,  \label{1}
\end{equation}
or, in other words: 
\begin{eqnarray}
&&\left\langle EL,el:\Lambda \right. \left| e_{1}l_{1},e_{2}l_{2}:\Lambda
\right\rangle _{d}  \notag \\
&=&\frac{1}{2\Lambda +1}\sum_{\lambda }\int \int d\mathbf{R}d\mathbf{r}%
\left\langle EL\left( \mathbf{R}\right) ,el\left( \mathbf{r}\right) :\Lambda
\lambda \right| \left. e_{1}l_{1}\left( \mathbf{r}_{1}\right)
,e_{2}l_{2}\left( \mathbf{r}_{2}\right) :\Lambda \lambda \right\rangle .
\label{2}
\end{eqnarray}
\label{2}Here $d$\ is a nonnegative real number (see Eq. (\ref{7})). The
two-particle wave functions with bound momenta are defined as: 
\begin{eqnarray}
\left| e_{1}l_{1}\left( \mathbf{r}_{1}\right) ,e_{2}l_{2}\left( \mathbf{r}%
_{2}\right) :\Lambda \lambda \right\rangle &\equiv &\left\{ \phi
_{e_{1}l_{1}}\left( \mathbf{r}_{1}\right) \otimes \phi _{e_{2}l_{2}}\left( 
\mathbf{r}_{2}\right) \right\} _{\Lambda \lambda }  \notag \\
&=&\sum_{m_{1}m_{2}}\left\langle l_{1}m_{1},l_{2}m_{2}\right. \left| \Lambda
\lambda \right\rangle \phi _{e_{1}l_{1}m_{1}}\left( \mathbf{r}_{1}\right)
\phi _{e_{2}l_{2}m_{2}}\left( \mathbf{r}_{2}\right)  \label{3}
\end{eqnarray}
and the properly orthonormalized harmonic-oscillator function is given by 
\begin{equation}
\phi _{elm}\left( \mathbf{r}\right) =\left( -1\right) ^{n}\left[ \frac{%
2\left( n!\right) }{\Gamma \left( n+l+3/2\right) }\right] ^{\frac{1}{2}}\exp
\left( -r^{2}/2\right) r^{l}L_{n}^{\left( l+1/2\right) }\left( r^{2}\right)
Y_{lm}\left( \mathbf{r/}r\right) ,  \label{4}
\end{equation}
where the\emph{\ }corresponding dimensionless\emph{\ }eigenvalue equals $%
\left( e+3/2\right) ,\quad $ $n=\left( e-l\right) /2=0,1,2,....$ We prefer
the\emph{\ }quantum number $e=2n+l,$ rather than $n,$\ due to conservation
of the total oscillator energy on both sides of the bracket: 
\begin{equation}
e_{1}+e_{2}=E+e.  \label{5}
\end{equation}
Obviously, this relation gives the requirement for the angular momenta: 
\begin{equation}
\left( -1\right) ^{l_{1}+l_{2}}=\left( -1\right) ^{L+l}.  \label{6}
\end{equation}
Let us next define the transformation of variables present in the expression
for the brackets in the following way: 
\begin{equation}
\left( 
\begin{tabular}{l}
$\mathbf{R}$ \\ 
$\mathbf{r}$%
\end{tabular}
\right) =\left( 
\begin{tabular}{ll}
$\sqrt{\frac{d}{1+d}}$ & $\sqrt{\frac{1}{1+d}}$ \\ 
$\sqrt{\frac{1}{1+d}}$ & $-\sqrt{\frac{d}{1+d}}$%
\end{tabular}
\right) \left( 
\begin{tabular}{l}
$\mathbf{r}_{1}$ \\ 
$\mathbf{r}_{2}$%
\end{tabular}
\right) ;\quad \left( 
\begin{tabular}{l}
$\mathbf{r}_{1}$ \\ 
$\mathbf{r}_{2}$%
\end{tabular}
\right) =\left( 
\begin{tabular}{ll}
$\sqrt{\frac{d}{1+d}}$ & $\sqrt{\frac{1}{1+d}}$ \\ 
$\sqrt{\frac{1}{1+d}}$ & $-\sqrt{\frac{d}{1+d}}$%
\end{tabular}
\right) \left( 
\begin{tabular}{l}
$\mathbf{R}$ \\ 
$\mathbf{r}$%
\end{tabular}
\right) .  \label{7}
\end{equation}
We prefer the above order for the variables and the coupling of the angular
momenta because this produces an orthogonal and, at the same time, symmetric
transformation matrix, hence, giving higher symmetry to the brackets: 
\begin{equation}
\left\langle EL,el:\Lambda \right. \left| e_{1}l_{1},e_{2}l_{2}:\Lambda
\right\rangle _{d}=\left\langle e_{1}l_{1},e_{2}l_{2}:\Lambda \right. \left|
EL,el:\Lambda \right\rangle _{d}.  \label{8}
\end{equation}
Any orthogonal matrix of second order can be represented in form (\ref{7}).
If problems arise, the matrix can be easily rewritten in this form. The
complete solution of this problem is as follows: In general, the matrix in (%
\ref{7}) has the form$\left( 
\begin{tabular}{ll}
$\sin \theta $ & $\cos \theta $ \\ 
$\cos \theta $ & $-\sin \theta $%
\end{tabular}
\right) $ and possesses the required distribution of signs, i.e., $\left( 
\begin{array}{cc}
+ & + \\ 
+ & -%
\end{array}
\right) $,\ only in the\emph{\ }case $0\leq \theta \leq \pi /2.$ When $\pi
/2\leq \theta \leq \pi ,$ the signs change so that the\emph{\ }matrix looks
like $\left( 
\begin{array}{cc}
+ & - \\ 
- & -%
\end{array}
\right) $,\ but by an elementary transformation of variables, namely $%
\mathbf{R\rightarrow -r}$ and $\mathbf{r\rightarrow R}$, we can transform it
to the previous form, i.e., only one minus sign in the\emph{\ }lower right
corner. If $\pi \leq \theta \leq 3\pi /2,$ distribution of signs is $\left( 
\begin{array}{cc}
- & - \\ 
- & +%
\end{array}
\right) $ and the necessary transformation of variables is $\mathbf{R}%
\rightarrow -\mathbf{R}$ and $\mathbf{r}\rightarrow -\mathbf{r}.$ The last
case, when $3\pi /2\leq \theta \leq 2\pi ,$ gives the matrix $\left( 
\begin{array}{cc}
- & + \\ 
+ & +%
\end{array}
\right) $ \emph{;} the correcting transformation is now $\mathbf{%
R\rightarrow r}$ and $\mathbf{r}\rightarrow \mathbf{-R}.$ When the
transformation is given by a\emph{\ }nonsymmetric orthogonal matrix, the
HOBs must, in general, be written in the\emph{\ }form 
\begin{equation}
\left\langle 
\begin{array}{c}
EL \\ 
\mathbf{R}%
\end{array}
, 
\begin{array}{c}
el \\ 
\mathbf{r}%
\end{array}
:\Lambda \right. \left| 
\begin{array}{c}
e_{1}l_{1} \\ 
\mathbf{r}_{1}%
\end{array}
, 
\begin{array}{c}
e_{2}l_{2} \\ 
\mathbf{r}_{2}%
\end{array}
:\Lambda \right\rangle _{d},  \label{9}
\end{equation}
which clearly demonstrates the correspondence between quantum numbers and
variables. Some authors define these brackets without referencing this
correspondence; in such a case, one must be careful, when employing some of
their symmetries.

The HOBs, defined as above, are the real entries of an orthogonal matrix.
Consequently, they obey usual orthogonality conditions: 
\begin{equation}
\sum_{EL,el}\left\langle e_{1}l_{1},e_{2}l_{2}:\Lambda \right. \left|
EL,el:\Lambda \right\rangle _{d}\left\langle EL,el:\Lambda \right. \left|
e_{1}^{\prime }l_{1}^{\prime },e_{2}^{\prime }l_{2}^{\prime }:\Lambda
\right\rangle _{d}=\delta _{e_{1}l_{1},e_{1}^{\prime }l_{1}^{\prime }}\delta
_{e_{2}l_{2},e_{2}^{\prime }l_{2}^{\prime }}.  \label{10}
\end{equation}
\begin{equation}
\sum_{e_{1}l_{1},e_{2}l_{2}}\left\langle EL,el:\Lambda \right. \left|
e_{1}l_{1},e_{2}l_{2}:\Lambda \right\rangle _{d}\left\langle
e_{1}l_{1},e_{2}l_{2}:\Lambda \right. \left| E^{\prime }L^{\prime
},e^{\prime }l^{\prime }:\Lambda \right\rangle _{d}=\delta _{EL,E^{\prime
}L^{\prime }}\delta _{el,e^{\prime }l^{\prime }}.  \label{11}
\end{equation}
Here and below we apply the compact expressions for the Kronecker deltas,
e.g., $\delta _{EL,E^{\prime }L^{\prime }}\equiv \delta _{E,E^{\prime
}}\delta _{L,L^{\prime }}.$ The symmetry properties of the coefficients are: 
\begin{align}
& \left\langle e_{1}l_{1},e_{2}l_{2}:\Lambda \right. \left| EL,el:\Lambda
\right\rangle _{d}  \notag \\
& =\left\langle EL,el:\Lambda \right. \left| e_{1}l_{1},e_{2}l_{2}:\Lambda
\right\rangle _{d}  \label{12.1} \\
& =\left( -1\right) ^{L+l_{2}}\left\langle e_{2}l_{2},e_{1}l_{1}:\Lambda
\right. \left| el,EL:\Lambda \right\rangle _{d}  \label{12.2} \\
& =\left( -1\right) ^{\Lambda -L}\left\langle e_{2}l_{2},e_{1}l_{1}:\Lambda
\right. \left| EL,el:\Lambda \right\rangle _{1/d}  \label{12.3} \\
& =\left( -1\right) ^{\Lambda -l_{1}}\left\langle
e_{1}l_{1},e_{2}l_{2}:\Lambda \right. \left| el,EL:\Lambda \right\rangle
_{1/d}.  \label{12.4}
\end{align}
The symmetry defined in (\ref{12.1}) follows from the definition of the
coefficients and the requirement for the angular momenta (\ref{6}); it is
already given in (\ref{8}). The next symmetry, (\ref{12.2}), is not so
trivial. It can be derived using the fact that the same transformation
matrix connects not only the original, but also the transformed variables: 
\begin{equation}
\left( 
\begin{tabular}{l}
$-\mathbf{r}$ \\ 
$\mathbf{R}$%
\end{tabular}
\right) =\left( 
\begin{tabular}{ll}
$\sqrt{\frac{d}{1+d}}$ & $\sqrt{\frac{1}{1+d}}$ \\ 
$\sqrt{\frac{1}{1+d}}$ & $-\sqrt{\frac{d}{1+d}}$%
\end{tabular}
\right) \left( 
\begin{tabular}{l}
$\mathbf{r}_{2}$ \\ 
$-\mathbf{r}_{1}$%
\end{tabular}
\right) .  \label{13}
\end{equation}
Using this condition for corresponding wave functions, one immediately
obtains the symmetry relation (\ref{12.2}). The symmetries (\ref{12.3}) and (%
\ref{12.4}) are even more complicated. The first one, (\ref{12.3}), is based
on observation that 
\begin{equation}
\left( 
\begin{tabular}{l}
$\mathbf{R}$ \\ 
$-\mathbf{r}$%
\end{tabular}
\right) =\left( 
\begin{tabular}{ll}
$\sqrt{\frac{1}{1+d}}$ & $\sqrt{\frac{d}{1+d}}$ \\ 
$\sqrt{\frac{d}{1+d}}$ & $-\sqrt{\frac{1}{1+d}}$%
\end{tabular}
\right) \left( 
\begin{tabular}{l}
$\mathbf{r}_{2}$ \\ 
$\mathbf{r}_{1}$%
\end{tabular}
\right) ,  \label{14}
\end{equation}
where the\emph{\ }transformation matrix corresponds to the\emph{\ }parameter
value $1/d$ instead of the original value $d$. Let us illustrate the
derivation of the symmetry in this case. The expression for the\emph{\ }%
brackets can be represented in the form: 
\begin{align}
& \left\langle e_{1}l_{1},e_{2}l_{2}:\Lambda \right. \left| EL,el:\Lambda
\right\rangle _{d}  \notag \\
& =\frac{1}{2\Lambda +1}\sum_{\lambda }\left\langle e_{1}l_{1}\left( \mathbf{%
r}_{1}\right) ,e_{2}l_{2}\left( \mathbf{r}_{2}\right) :\Lambda \lambda
\right| \left. EL\left( \mathbf{R}\right) ,el\left( \mathbf{r}\right)
:\Lambda \lambda \right\rangle  \notag \\
& =\frac{1}{2\Lambda +1}\sum_{\lambda }\int \int d\mathbf{r}_{1}d\mathbf{r}%
_{2}\left\{ \phi _{e_{1}l_{1}}\left( \mathbf{r}_{1}\right) \otimes \phi
_{e_{2}l_{2}}\left( \mathbf{r}_{2}\right) \right\} _{\Lambda \lambda
}^{+}\left\{ \phi _{EL}\left( \mathbf{R}\right) \otimes \phi _{el}\left( 
\mathbf{r}\right) \right\} _{\Lambda \lambda }.  \label{15}
\end{align}
Now it is enough to reorder the variables according to both sides of (\ref%
{14}). To do this, the following simple, well-known expressions are
necessary: 
\begin{equation}
\phi _{elm}\left( -\mathbf{r}\right) =\left( -1\right) ^{l}\phi _{elm}\left( 
\mathbf{r}\right)  \label{16}
\end{equation}
and 
\begin{equation}
\left\{ \phi _{e_{1}l_{1}}\left( \mathbf{r}_{1}\right) \otimes \phi
_{e_{2}l_{2}}\left( \mathbf{r}_{2}\right) \right\} _{\Lambda \lambda
}=\left( -1\right) ^{l_{1}+l_{2}-\Lambda }\left\{ \phi _{e_{2}l_{2}}\left( 
\mathbf{r}_{2}\right) \otimes \phi _{e_{1}l_{1}}\left( \mathbf{r}_{1}\right)
\right\} _{\Lambda \lambda }.  \label{17}
\end{equation}
Applying these expressions, the right-hand side of (\ref{15}) can be
rewritten as 
\begin{align}
& \left\langle e_{1}l_{1},e_{2}l_{2}:\Lambda \right. \left| EL,el:\Lambda
\right\rangle _{d}  \notag \\
& =\left( -1\right) ^{l_{1}+l_{2}-\Lambda +l}\frac{1}{2\Lambda +1}%
\sum_{\lambda }\int \int d\mathbf{r}_{1}d\mathbf{r}_{2}\left\{ \phi
_{e_{2}l_{2}}\left( \mathbf{r}_{2}\right) \otimes \phi _{e_{1}l_{1}}\left( 
\mathbf{r}_{1}\right) \right\} _{\Lambda \lambda }^{+}\left\{ \phi
_{EL}\left( \mathbf{R}\right) \otimes \phi _{el}\left( -\mathbf{r}\right)
\right\} _{\Lambda \lambda }  \notag \\
& =\left( -1\right) ^{l+L-\Lambda +l}\left\langle
e_{2}l_{2},e_{1}l_{1}:\Lambda \right. \left| EL,el:\Lambda \right\rangle
_{1/d}  \notag \\
& =\left( -1\right) ^{\Lambda -L}\left\langle e_{2}l_{2},e_{1}l_{1}:\Lambda
\right. \left| EL,el:\Lambda \right\rangle _{1/d}.  \label{18}
\end{align}
The last symmetry, (\ref{12.4}), follows from the transformation 
\begin{equation}
\left( 
\begin{tabular}{l}
$\mathbf{r}$ \\ 
$\mathbf{R}$%
\end{tabular}
\right) =\left( 
\begin{tabular}{ll}
$\sqrt{\frac{1}{1+d}}$ & $\sqrt{\frac{d}{1+d}}$ \\ 
$\sqrt{\frac{d}{1+d}}$ & $-\sqrt{\frac{1}{1+d}}$%
\end{tabular}
\right) \left( 
\begin{tabular}{l}
$\mathbf{r}_{1}$ \\ 
$-\mathbf{r}_{2}$%
\end{tabular}
\right) .  \label{19}
\end{equation}
The self-consistence of the transformations and, hence, the symmetries can
be checked by applying two transformations, say (\ref{12.3}) and (\ref{12.4}%
), for the same bracket. The result yields an identity, as expected: 
\begin{align}
\left\langle e_{1}l_{1},e_{2}l_{2}:\Lambda \right. \left| EL,el:\Lambda
\right\rangle _{d}& =\left( -1\right) ^{\Lambda -L}\left\langle
e_{2}l_{2},e_{1}l_{1}:\Lambda \right. \left| EL,el:\Lambda \right\rangle
_{1/d}  \notag \\
& =\left( -1\right) ^{\Lambda -L}\left( -1\right) ^{\Lambda
-l_{2}}\left\langle e_{2}l_{2},e_{1}l_{1}:\Lambda \right. \left|
el,EL:\Lambda \right\rangle _{d}  \notag \\
& =\left\langle e_{1}l_{1},e_{2}l_{2}:\Lambda \right. \left| EL,el:\Lambda
\right\rangle _{d}.  \label{20}
\end{align}
Hence, our definition of brackets is based on an assumption that variables
in brackets are arranged in a fixed way, and the\emph{\ }corresponding
transformation matrix is given as in (\ref{7}): 
\begin{equation}
\left\langle \mathbf{R},\mathbf{r}\right. \left| \mathbf{r}_{1},\mathbf{r}%
_{2}\right\rangle _{d}=\left( 
\begin{tabular}{ll}
$\sqrt{\frac{d}{1+d}}$ & $\sqrt{\frac{1}{1+d}}$ \\ 
$\sqrt{\frac{1}{1+d}}$ & $-\sqrt{\frac{d}{1+d}}$%
\end{tabular}
\right) .  \label{21}
\end{equation}

\section{Compact expression}

The HOBs have been considered previously by a number of authors; however,
these results were quite complicated and led to expressions, whose
structures are not transparent. In our opinion, the simplest known
expression for the general oscillator bracket is the one derived by B.Buck
and A.C.Merchant in Ref.\cite{BM}: 
\begin{align}
& \left\langle EL,el:\Lambda \right. \left| e_{1}l_{1},e_{2}l_{2}:\Lambda
\right\rangle _{d}  \notag \\
& =i^{-\left( l_{1}+l_{2}+L+l\right) }\times 2^{-\left(
l_{1}+l_{2}+L+l\right) /4}  \notag \\
& \times \sqrt{\left( n_{1}\right) !\left( n_{2}\right) !\left( N\right)
!\left( n\right) !\left[ 2\left( n_{1}+l_{1}\right) +1\right] !!\left[
2\left( n_{2}+l_{2}\right) +1\right] !!\left[ 2\left( N+L\right) +1\right] !!%
\left[ 2\left( n+l\right) +1\right] !!}  \notag \\
& \times \sum_{abcdl_{a}l_{b}l_{c}l_{d}}\left( -1\right)
^{l_{a}+l_{b}+l_{c}}2^{\left( l_{a}+l_{b}+l_{c}+l_{d}\right) /2}d^{\left(
2a+l_{a}+2d+l_{d}\right) /2}\left( 1+d\right) ^{-\left(
2a+l_{a}+2b+l_{b}+2c+l_{c}+2d+l_{d}\right) /2}  \notag \\
& \times \frac{\left[ \left( 2l_{a}+1\right) \left( 2l_{b}+1\right) \left(
2l_{c}+1\right) \left( 2l_{d}+1\right) \right] }{a!b!c!d!\left[ 2\left(
a+l_{a}\right) +1\right] !!\left[ 2\left( b+l_{b}\right) +1\right] !!\left[
2\left( c+l_{c}\right) +1\right] !!\left[ 2\left( d+l_{d}\right) +1\right] !!%
}  \notag \\
& \times \left\{ 
\begin{array}{ccc}
l_{a} & l_{b} & l_{1} \\ 
l_{c} & l_{d} & l_{2} \\ 
L & l & \Lambda%
\end{array}
\right\} \left\langle l_{a}0l_{c}0\right| \left. L0\right\rangle
\left\langle l_{b}0l_{d}0\right| \left. l0\right\rangle \left\langle
l_{a}0l_{b}0\right| \left. l_{1}0\right\rangle \left\langle
l_{c}0l_{d}0\right| \left. l_{2}0\right\rangle ,  \label{22}
\end{align}
where $N=\left( E-L\right) /2,\quad n=\left( e-l\right) /2,\quad
n_{1}=\left( e_{1}-l_{1}\right) /2,$ \ and$\quad n_{2}=\left(
e_{2}-l_{2}\right) /2.$ This expression for the HOBs is derived using
harmonic-oscillator wave functions without the\emph{\ }phase multiplier $%
\left( -1\right) ^{n}$ present in our Eq. (\ref{4}). The introduction of
this modification results in a slightly different phase in the expression
for HOBs. This phase equals $\left( -1\right) ^{N+n+n_{1}+n_{2}}\equiv
\left( -1\right) ^{\left( L+l-l_{1}-l_{2}\right) /2}$.

Although the above summation runs over eight indices, the real summation is
only over five of them due to three independent constraints. This
constrained summation is similar to other, known, modern expressions for the
HOBs. Formula (\ref{22}) is very symmetric, and due to this symmetry, can be
rewritten in the following way: 
\begin{align}
& \left\langle EL,el:\Lambda \right. \left| e_{1}l_{1},e_{2}l_{2}:\Lambda
\right\rangle _{d}  \notag \\
& =d^{\left( e_{1}-e\right) /2}\left( 1+d\right) ^{-\left(
e_{1}+e_{2}\right) /2}\sum_{e_{a}l_{a}e_{b}l_{b}e_{c}l_{c}e_{d}l_{d}}\left(
-d\right) ^{e_{d}}\left\{ 
\begin{array}{ccc}
l_{a} & l_{b} & l_{1} \\ 
l_{c} & l_{d} & l_{2} \\ 
L & l & \Lambda%
\end{array}
\right\}  \notag \\
& \times G\left( e_{1}l_{1};e_{a}l_{a},e_{b}l_{b}\right) G\left(
e_{2}l_{2};e_{c}l_{c},e_{d}l_{d}\right) G\left(
EL;e_{a}l_{a},e_{c}l_{c}\right) G\left( el;e_{b}l_{b},e_{d}l_{d}\right) ,
\label{23}
\end{align}
where 
\begin{align}
G\left( e_{1}l_{1};e_{a}l_{a},e_{b}l_{b}\right) & =\sqrt{\left(
2l_{a}+1\right) \left( 2l_{b}+1\right) }\left\langle l_{a}0l_{b}0\right|
\left. l_{1}0\right\rangle  \notag \\
& \times \left[ \left( 
\begin{array}{c}
e_{1}-l_{1} \\ 
e_{a}-l_{a};e_{b}-l_{b}%
\end{array}
\right) \left( 
\begin{array}{c}
e_{1}+l_{1}+1 \\ 
e_{a}+l_{a}+1;e_{b}+l_{b}+1%
\end{array}
\right) \right] ^{1/2}  \label{24}
\end{align}
and the coefficients 
\begin{equation}
\left( 
\begin{array}{c}
n_{1} \\ 
n_{a};n_{b}%
\end{array}
\right) =\frac{\left( n_{1}\right) !!}{\left( n_{a}\right) !!\left(
n_{b}\right) !!}  \label{25}
\end{equation}
are trinomials defined for the parameter values $n_{1}\geq n_{a},n_{b}$ .
Moreover, the three parameters of these coefficients are all even or all odd
at the same time. It should be noted that the well-known binomial
coefficients can be expanded in trinomials, e.g., 
\begin{equation}
\left( 
\begin{array}{c}
n \\ 
k%
\end{array}
\right) =\left( 
\begin{array}{c}
2n \\ 
2k;2\left( n-k\right)%
\end{array}
\right) .  \label{26}
\end{equation}
In expression (\ref{23})\ we again have eight summation indices connected in
four pairs, $e_{\alpha }l_{\alpha }$ $\left( \alpha =a,b,c,d\right) $, with
the well-known relation between the\emph{\ }oscillator energy and the\emph{\ 
}angular\emph{\ }momentum: $e_{\alpha }$ is a nonnegative integer and $%
l_{\alpha }=$ $e_{\alpha },e_{\alpha }-2,...,1$ or $0.$ The constraints
amongst the energies are: 
\begin{eqnarray}
e_{a}+e_{c} &=&E,\quad e_{b}+e_{d}=e,  \notag \\
e_{a}+e_{b} &=&e_{1},\quad e_{c}+e_{d}=e_{2}.  \label{27}
\end{eqnarray}
Due to relation (\ref{5}), $\left( E+e=e_{1}+e_{2}\right) ,$ only three of
them are independent. The best choice for the independent summation indices
is $e_{d}$ and the four angular momenta $l_{\alpha }\quad \left( \alpha
=a,b,c,d\right) .$

\section{Sums of products}

The sums of HOBs occur when antisymmetrizing the translationally invariant
wave function, i.e., the function of Jacobi and intrinsic (spin, isospin,
etc.) variables, because any permutation of one-particle coordinates results
in orthogonal transformations of \ the Jacobian coordinates. These
permutations result in different orthogonal transformations; therefore,
according to (\ref{7}), in transformations with different values of \ the
parameters $d$. The keys to the derivation of the expressions for the sums
of products are Eq. (\ref{15}) and the possibility to represent any
orthogonal transformation in the form (\ref{21}). To illustrate this
procedure, let us take two expressions for the HOBs, as given by Eq. (\ref{1}%
), 
\begin{equation}
\left\{ \phi _{EL}\left( \mathbf{R}\right) \otimes \phi _{el}\left( \mathbf{r%
}\right) \right\} _{\Lambda \lambda
}=\sum_{e_{1}l_{1},e_{2}l_{2}}\left\langle EL,el:\Lambda \right. \left|
e_{1}l_{1},e_{2}l_{2}:\Lambda \right\rangle _{d}\left\{ \phi
_{e_{1}l_{1}}\left( \mathbf{r}_{1}\right) \otimes \phi _{e_{2}l_{2}}\left( 
\mathbf{r}_{2}\right) \right\} _{\Lambda \lambda },  \label{28}
\end{equation}%
with different variables and different orthogonal transformations, and
multiply them, the left side to the left side and the right side to the
right side. Next let us sum over the projection of the angular momentum and
integrate over both coordinates to obtain: 
\begin{align}
& \frac{1}{2\Lambda +1}\sum_{\lambda }\int \int d\mathbf{R}d\mathbf{r}%
\left\{ \phi _{EL}\left( \mathbf{R}\right) \otimes \phi _{el}\left( \mathbf{r%
}\right) \right\} _{\Lambda \lambda }^{+}\left\{ \phi _{E^{\prime }L^{\prime
}}\left( \mathbf{R}^{\prime }\right) \otimes \phi _{e^{\prime }l^{\prime
}}\left( \mathbf{r}^{\prime }\right) \right\} _{\Lambda \lambda }  \notag \\
& =\sum_{e_{1}l_{1}e_{2}l_{2}e_{1}^{\prime }l_{1}^{\prime }e_{2}^{\prime
}l_{2}^{\prime }}\left\langle EL,el:\Lambda \right. \left|
e_{1}l_{1},e_{2}l_{2}:\Lambda \right\rangle _{d}\left\langle E^{\prime
}L^{\prime },e^{\prime }l^{\prime }:\Lambda \right. \left| e_{1}^{\prime
}l_{1}^{\prime },e_{2}^{\prime }l_{2}^{\prime }:\Lambda \right\rangle
_{d^{\prime }}  \notag \\
& \times \frac{1}{2\Lambda +1}\sum_{\lambda }\int \int d\mathbf{r}_{1}d%
\mathbf{r}_{2}\left\{ \phi _{e_{1}l_{1}}\left( \mathbf{r}_{1}\right) \otimes
\phi _{e_{2}l_{2}}\left( \mathbf{r}_{2}\right) \right\} _{\Lambda \lambda
}^{+}\left\{ \phi _{e_{1}^{\prime }l_{1}^{\prime }}\left( \mathbf{r}%
_{1}^{\prime }\right) \otimes \phi _{e_{2}^{\prime }l_{2}^{\prime }}\left( 
\mathbf{r}_{2}^{\prime }\right) \right\} _{\Lambda \lambda }.  \label{29}
\end{align}%
Obviously, the variables $\mathbf{R},\mathbf{r}$ $\ $and $\mathbf{r}_{1},%
\mathbf{r}_{2}$ are connected by the orthogonal transformation$\
\left\langle \mathbf{R},\mathbf{r}\right. \left| \mathbf{r}_{1},\mathbf{r}%
_{2}\right\rangle _{d}$, Eq. (\ref{21}), and variables $\mathbf{R}^{\prime },%
\mathbf{r}^{\prime }$ $\ $and $\mathbf{r}_{1}^{\prime },\mathbf{r}%
_{2}^{\prime }$ are connected by another orthogonal transformation$\
\left\langle \mathbf{R}^{\prime },\mathbf{r}^{\prime }\right. \left| \mathbf{%
r}_{1}^{\prime },\mathbf{r}_{2}^{\prime }\right\rangle _{d^{\prime }}$ of
the same structure. In the\emph{\ }case when $\mathbf{r}_{1},\mathbf{r}_{2}$
and $\mathbf{r}_{1}^{\prime },\mathbf{r}_{2}^{\prime }$\ are also connected
by some orthogonal transformation, say $\left\langle \mathbf{r}_{1},\mathbf{r%
}_{2}\right. \left| \mathbf{r}_{1}^{\prime },\mathbf{r}_{2}^{\prime
}\right\rangle _{d_{0}}$, both integrals on the\emph{\ }left and the\emph{\ }%
right sides of Eq. (\ref{29}) are expressible as HOBs. Thus, Eq. (\ref{29})
can be rewritten as 
\begin{align}
& \left\langle EL,el:\Lambda \right. \left| E^{\prime }L^{\prime },e^{\prime
}l^{\prime }:\Lambda \right\rangle _{D}  \notag \\
& =\sum_{e_{1}l_{1}e_{2}l_{2}e_{1}^{\prime }l_{1}^{\prime }e_{2}^{\prime
}l_{2}^{\prime }}\left\langle EL,el:\Lambda \right. \left|
e_{1}l_{1},e_{2}l_{2}:\Lambda \right\rangle _{d}\left\langle
e_{1}l_{1},e_{2}l_{2}:\Lambda \right. \left| e_{1}^{\prime }l_{1}^{\prime
},e_{2}^{\prime }l_{2}^{\prime }:\Lambda \right\rangle _{d_{0}}  \notag \\
& \times \left\langle e_{1}^{\prime }l_{1}^{\prime },e_{2}^{\prime
}l_{2}^{\prime }:\Lambda \right. \left| E^{\prime }L^{\prime },e^{\prime
}l^{\prime }:\Lambda \right\rangle _{d^{\prime }}.  \label{30}
\end{align}%
The only remaining problem is the definition of the\emph{\ }parameter $D$.
Having in mind that all transformations of variables are orthogonal and
well-defined, we can expand 
\begin{equation}
\left( 
\begin{tabular}{l}
$\mathbf{R}$ \\ 
$\mathbf{r}$%
\end{tabular}%
\ \right) =\left( 
\begin{tabular}{ll}
$\sqrt{\frac{d}{1+d}}$ & $\sqrt{\frac{1}{1+d}}$ \\ 
$\sqrt{\frac{1}{1+d}}$ & $-\sqrt{\frac{d}{1+d}}$%
\end{tabular}%
\ \right) \left( 
\begin{tabular}{ll}
$\sqrt{\frac{d_{0}}{1+d_{0}}}$ & $\sqrt{\frac{1}{1+d_{0}}}$ \\ 
$\sqrt{\frac{1}{1+d_{0}}}$ & $-\sqrt{\frac{d_{0}}{1+d_{0}}}$%
\end{tabular}%
\ \right) \left( 
\begin{tabular}{ll}
$\sqrt{\frac{d^{\prime }}{1+d^{\prime }}}$ & $\sqrt{\frac{1}{1+d^{\prime }}}$
\\ 
$\sqrt{\frac{1}{1+d^{\prime }}}$ & $-\sqrt{\frac{d}{1+d^{\prime }}}$%
\end{tabular}%
\ \right) \left( 
\begin{tabular}{l}
$\mathbf{R}^{\prime }$ \\ 
$\mathbf{r}^{\prime }$%
\end{tabular}%
\ \right) .  \label{31}
\end{equation}%
Multiplying the above matrices, one obtains the form: 
\begin{equation}
\left( 
\begin{tabular}{l}
$\mathbf{R}$ \\ 
$\mathbf{r}$%
\end{tabular}%
\ \right) =\left( 
\begin{tabular}{ll}
$\sqrt{\frac{D}{1+D}}$ & $\sqrt{\frac{1}{1+D}}$ \\ 
$\sqrt{\frac{1}{1+D}}$ & $-\sqrt{\frac{D}{1+D}}$%
\end{tabular}%
\ \right) \left( 
\begin{tabular}{l}
$\mathbf{R}^{\prime }$ \\ 
$\mathbf{r}^{\prime }$%
\end{tabular}%
\ \right) ,  \label{32}
\end{equation}%
where 
\begin{equation}
\sqrt{D}=\frac{\sqrt{d}+\sqrt{d^{\prime }}-\sqrt{d_{0}}\left( 1-\sqrt{%
dd^{\prime }}\right) }{1-\sqrt{dd^{\prime }}+\sqrt{d_{0}}\left( \sqrt{d}+%
\sqrt{d^{\prime }}\right) },  \label{33}
\end{equation}%
and restrictions for values of the\emph{\ }parameter $d_{0}$ are as follows:
If $dd^{\prime }\leq 1,$ then\ $\sqrt{d_{0}}\leq \left( \sqrt{d}+\sqrt{%
d^{\prime }}\right) /\left( 1-\sqrt{dd^{\prime }}\right) $; if $dd^{\prime
}\geq 1,$ then $\sqrt{d_{0}}\geq \left( \sqrt{dd^{\prime }}-1\right) /\left( 
\sqrt{d}+\sqrt{d^{\prime }}\right) $. The sum in Eq. (\ref{31}) of the
product of three HOBs with different values of the parameter $d$\ can be
simplified to sums of products of \ only two HOBs. To do this, the values of
the brackets corresponding to the extreme values of the parameter $d$ are
necessary, i.e., $d=0$\ or $d\rightarrow \infty .$\ The transformations of
variables in these cases are trivial and HOBs can be calculated directly
from definition (\ref{15}), yielding: 
\begin{align}
\left\langle EL,el:\Lambda \right. \left| e_{1}l_{1},e_{2}l_{2}:\Lambda
\right\rangle _{d=0}& =\left( -1\right) ^{L+l-\Lambda }\delta
_{e_{1}l_{1},el}\delta _{e_{2}l_{2},EL},  \notag \\
\left\langle EL,el:\Lambda \right. \left| e_{1}l_{1},e_{2}l_{2}:\Lambda
\right\rangle _{d\rightarrow \infty }& =\left( -1\right) ^{l}\delta
_{e_{1}l_{1},EL}\delta _{e_{2}l_{2},el}.  \label{34}
\end{align}%
Using the HOB with the parameter $d_{0}\rightarrow \infty ,$\ one obtains
the following expressions (the value $d_{0}\rightarrow \infty $\ is
consistent only with $dd^{\prime }\geq 1$)\ : 
\begin{align}
& \left\langle EL,el:\Lambda \right. \left| E^{\prime }L^{\prime },e^{\prime
}l^{\prime }:\Lambda \right\rangle _{\left( \left( \sqrt{dd^{\prime }}%
-1\right) /\left( \sqrt{d}+\sqrt{d^{\prime }}\right) \right) ^{2}}  \notag \\
& =\sum_{e_{1}l_{1},e_{2}l_{2}}\left( -1\right) ^{l_{2}}\left\langle
EL,el:\Lambda \right. \left| e_{1}l_{1},e_{2}l_{2}:\Lambda \right\rangle
_{d}\left\langle e_{1}l_{1},e_{2}l_{2}:\Lambda \right. \left| E^{\prime
}L^{\prime },e^{\prime }l^{\prime }:\Lambda \right\rangle _{d^{\prime }}.
\label{35}
\end{align}%
Using the HOB with the parameter $d_{0}=0$ (consistent only with $dd^{\prime
}\leq 1$),\ one obtains the result: 
\begin{align}
& \left\langle EL,el:\Lambda \right. \left| E^{\prime }L^{\prime },e^{\prime
}l^{\prime }:\Lambda \right\rangle _{\left( \left( \sqrt{d}+\sqrt{d^{\prime }%
}\right) /\left( 1-\sqrt{dd^{\prime }}\right) \right) ^{2}}  \notag \\
& =\left( -1\right) ^{L^{\prime }+\Lambda
}\sum_{e_{1}l_{1},e_{2}l_{2}}\left( -1\right) ^{l_{2}}\left\langle
EL,el:\Lambda \right. \left| e_{1}l_{1},e_{2}l_{2}:\Lambda \right\rangle
_{d}\left\langle e_{1}l_{1},e_{2}l_{2}:\Lambda \right. \left| e^{\prime
}l^{\prime },E^{\prime }L^{\prime }:\Lambda \right\rangle _{d^{\prime }}.
\label{36}
\end{align}%
These sums appear, when calculating the matrix element of the permutation
operator of the particle coordinates in the translationally invariant basis
of oscillator functions, see \cite{NKB}. They are consistent: at $d^{\prime
}=1/d$ in both cases one can easily obtain the normalization condition for
HOBs.

\section{Fortran code}

Using the above results, we have also developed a Fortran code for the fast
and precise calculation of the HOBs in large quantities. Such a computer
code is needed because any nuclear calculation, including mentioned above
antisymmetrization of \ a translationally invariant basis, requires large
matrices. As one can see from the expression for HOBs, Eqs. (\ref{23}) - (%
\ref{25}), the main elements of these expressions are the Clebsch-Gordan
coefficients with zero angular momenta projections, the $9-j$ symbols and
trinomial coefficients. Our code is based on the observation that all
group-theoretical expressions can be represented as products or sums of
products of binomial coefficients. Binomial coefficients are far more
acceptable for large calculations than are factorials. For example, to
represent $50!$ exactly requires a mantissa with $53$ significant numbers.
At the same time, the mantissa of the binomial $\binom{50}{25}$ requires
only $15$ significant numbers, hence it can be stored using real numbers of
double precision. Having this in mind, for Clebsch-Gordan coefficients we
use the expression: 
\begin{align}
& \left\langle l_{1}0l_{2}0\right| \left. l0\right\rangle   \notag \\
& =\left( -1\right) ^{\left( l_{1}+l_{2}-l\right) /2}\left[ \binom{2l}{%
l_{1}-l_{2}+l}\binom{l_{1}+l_{2}+l+1}{2l+1}\right] ^{-1/2}\binom{l}{\left(
l_{1}-l_{2}+l\right) /2}\binom{\left( l_{1}+l_{2}+l\right) /2}{l}.
\label{38}
\end{align}%
For the $9-j$ symbols we first express them in terms of $6-j$ symbols,\emph{%
\ }using the well-known formula: 
\begin{equation}
\left\{ 
\begin{array}{ccc}
j_{1} & j_{2} & j_{3} \\ 
l_{1} & l_{2} & l_{3} \\ 
k_{1} & k_{2} & k_{3}%
\end{array}%
\right\} =\sum_{x}\left( -1\right) ^{2x}\left( 2x+1\right) \left\{ 
\begin{array}{ccc}
j_{1} & j_{2} & j_{3} \\ 
l_{3} & k_{3} & x%
\end{array}%
\right\} \left\{ 
\begin{array}{ccc}
l_{1} & l_{2} & l_{3} \\ 
j_{2} & x & k_{2}%
\end{array}%
\right\} \left\{ 
\begin{array}{ccc}
k_{1} & k_{2} & k_{3} \\ 
x & j_{1} & l_{1}%
\end{array}%
\right\} ,  \label{39}
\end{equation}%
and then utilize the $6-j$ expression in terms of binomials: 
\begin{align}
\left\{ 
\begin{array}{ccc}
a & b & e \\ 
d & c & f%
\end{array}%
\right\} & =\frac{\left( -1\right) ^{a+d+e+f}}{a+c-f+1}\left[ \frac{\left(
a+b+e+1\right) \left( c+d+e+1\right) }{\left( a+c+f+1\right) \left(
b+d+f+1\right) }\right] ^{1/2}  \notag \\
& \times \left[ \frac{\binom{a+c+f}{2a}\binom{2a}{a+c-f}\binom{c+d+e}{2d}%
\binom{2d}{c+d-e}\binom{a+e+b}{2e}\binom{2e}{a+e-b}}{\binom{b+f+d}{2f}\binom{%
2f}{b+f-d}}\right] ^{1/2}  \notag \\
& \times \sum_{z}\left( -1\right) ^{z}\frac{\binom{a+f-c}{z}\binom{f+c-a}{%
d+f-b-z}}{\binom{a+c+f}{a+d+f-e-z}\binom{b+c+e-f+1+z}{c+a-f+1}}.  \label{40}
\end{align}%
This formula for the $6-j$ symbol is well-balanced because every unit summed
over has an equal number of binomials in the numerator and in the
denominator. This results in the highest precision for the sum. Our code
starts the calculations of the HOBs by filling the arrays of \ the binomial
and trinomial coefficients. For the binomials we use recurrence formulas
with corrections taking into account the fact that the binomial coefficients
are integer numbers. For the three-dimensional array of trinomials we apply
a completely analogous method. When both arrays are filled, the calculation
of the HOBs is an extremely fast and precise operation. In Table 1 we give
values of 
\begin{equation}
\delta \left( E_{0}\right) =\sum_{0}^{E_{0}}\left|
\sum_{e_{1}l_{1},e_{2}l_{2}}\left\langle EL,el:\Lambda \right. \left|
e_{1}l_{1},e_{2}l_{2}:\Lambda \right\rangle _{d}\left\langle
e_{1}l_{1},e_{2}l_{2}:\Lambda \right. \left| E^{\prime }L^{\prime
},e^{\prime }l^{\prime }:\Lambda \right\rangle _{d}-\delta _{EL,E^{\prime
}L^{\prime }}\delta _{el,e^{\prime }l^{\prime }}\right| ,  \label{41}
\end{equation}%
which characterize the precision of our calculations. Equation (\ref{41})
represents the sum of the absolute values of the deviations of the
calculated normalization conditions for the HOBs from the exact values given
by Eq. (\ref{11}). The first sum runs over all free parameters of the
normalization condition, i.e., $E,L,e,l,E^{\prime },L^{\prime },e^{\prime
},l^{\prime }$ and $\Lambda $, taking all allowed values between $0$\ and $%
E_{0}.$ The total number of HOBs calculated for a given value of $E_{0}$\
equals $N\left( E_{0}\right) .$ The processor time (in seconds), necessary
to perform these calculations on a personal computer, is listed under $%
T\left( E_{0}\right) .$

Table 1

$ 
\begin{tabular}{|l|l|l|l|}
\hline
$E_{0}$ & $N\left( E_{0}\right) $ & $\delta \left( E_{0}\right) $ & $T\left(
E_{0}\right) $ \\ \hline
$3$ & $4,672$ & $0.2956\ast 10^{-14}$ & $0.28$ \\ \hline
$4$ & $32,054$ & $0.6832\ast 10^{-13}$ & $1.65$ \\ \hline
$5$ & $157,648$ & $0.5028\ast 10^{-12}$ & $11.9$ \\ \hline
$6$ & $658,000$ & $0.2394\ast 10^{-11}$ & $74.7$ \\ \hline
$7$ & $2,298,144$ & $0.7336\ast 10^{-10}$ & $385$ \\ \hline
$8$ & $7,165,706$ & $0.7032\ast 10^{-9}$ & $1,760$ \\ \hline
$9$ & $19,973,174$ & $0.2057\ast 10^{-7}$ & $6,830$ \\ \hline
$10$ & $51,349,192$ & $0.1306\ast 10^{-6}$ & $25,100$ \\ \hline
$11$ & $122,193,968$ & $0.2131\ast 10^{-5}$ & $79,900$ \\ \hline
$12$ & $273,872,766$ & $0.3481\ast 10^{-4}$ & $248,000$ \\ \hline
\end{tabular}
$

As Table 1 clearly demonstrates, our method for calculating the HOBs is fast
and universal and, hence, applicable to any calculations using a basis of
many-particle harmonic-oscillator functions. Our Fortran code is available
for general use and distribution, as described in the Conclusions.

\section{Conclusions}

Starting with the result of Ref.\cite{BM} for the General
Harmonic-Oscillator Brackets (HOBs), we have simplified this earlier
expression into a more compact and highly symmetrical relationship, given in
Eqs. (\ref{23})-(\ref{24}). Our new result for the HOBs is well-suited for
fast and precise numerical calculations of large quantities of HOBs. The
numerical procedure is based upon independent calculations of the
Clebsch-Gordan coefficients of zero angular momentum projections, the $6-j$
symbols and the $9-j$ symbols using binomial and ''trinomial'' coefficients,
which are precalculated and stored in appropriate arrays. The Fortran code,
which we have written for computing the HOBs, is available for general use
and can be obtained from the Vytautas Magnus University, Kaunas Lithuania,
web server at\emph{\ }http://www.nuclear.physics.vdu.lt/.


\begin{thebibliography}{99}
\bibitem{NS69} Neudachin V.G., and Smirnov Ju.F., Nucleonic associations in
light atomic nuclei, Nauka, Moscow 1969, 414 p. (in Russian).

\bibitem{VV70} Vanagas V., Algebraic methods in nuclear theory, Mintis,
Vilnius, 1971, 378 p. (in Russian).

\bibitem{GK89} Kamuntavi\v{c}ius G.P., Sov. Journ. Part. and Nuclei, \textbf{%
20}, N2, 261 (1989).

\bibitem{NKB} Navratil P., Kamuntavi\v{c}ius G.P., Barrett B.R., Phys. Rev.
C, \textbf{61}, 044001 (2000).

\bibitem{SS77} Smirnov Yu.F., Shitikova K.V., Sov. Journ. Part. and Nuclei, 
\textbf{8}, 847 (1977).

\bibitem{Talmi} Talmi I., Helv. Phys. Acta, \textbf{25}, 185 (1952).

\bibitem{Mosh} Moshinsky M., Nucl. Phys., \textbf{13}, 104 (1959).

\bibitem{Smirn} Smirnov Yu. F., Nucl. Phys., \textbf{27}, 177 (1961); 
\textbf{39}, 346 (1962).

\bibitem{Trlifai} Trlifai L., Phys. Rev. C, \textbf{5}, 1534 (1972).

\bibitem{BM} Buck B., Merchant A.C., Nucl. Phys. \textbf{A600}, 387 (1996).
\end{thebibliography}
\end{document}